\documentclass[a4paper,fleqn,twoside]{article}
\usepackage{espcrc2}
\usepackage{mathbbol}

\usepackage{epsfig}

\def\lsi{\raise0.3ex\hbox{$<$\kern-0.75em\raise-1.1ex\hbox{$\sim$}}}
\def\gsi{\raise0.3ex\hbox{$>$\kern-0.75em\raise-1.1ex\hbox{$\sim$}}}
\newcommand{\lsim}{\mathop{\lsi}}

\title{Testing a Topology Conserving Gauge Action in Lattice QCD}
\author{S. Shcheredin \address[HU]{Institut f\"ur Physik, Humboldt 
Universit\"at zu Berlin, Newtonstr.\ 15, 12489 Berlin, Germany},
W. Bietenholz \addressmark[HU],
K. Jansen \address[DESY]{NIC/DESY Zeuthen, Platanenallee 6, 
D-15738 Zeuthen, Germany},
K.-I. Nagai \addressmark[DESY],
S. Necco \addressmark[Mars]\address[Mars]{Centre de Physique 
Th\'{e}orique, Luminy, Case 907, 
F-13288 Marseille Cedex 9, France} and
L. Scorzato \addressmark[HU]}

\begin{document}


\begin{abstract}

\vspace{-2mm}

We study lattice QCD with a gauge action, which suppresses small plaquette 
values. Thus the MC history is confined to a single topological sector
over a significant time, while other observables are decorrelated.
This enables the cumulation of statistics with 
a specific topological charge, which is needed for simulations of QCD in the 
$\epsilon$-regime. The same action may also be useful
for simulations with dynamical quarks. The update is performed with a 
local HMC algorithm.

\vspace{-6mm}

\end{abstract}

\vspace{-11mm}

\maketitle

\section{MOTIVATION}

\vspace{-2mm}

Our goal is to explore the applicability of lattice gauge actions
for QCD, which are designed such that small plaquette values are
strongly suppressed. If such an action can be identified, we expect
the following virtues:
\begin{itemize}
\vspace{-1mm}
\item an acceleration of dynamical fermion simulations
\vspace{-1mm}
\item control over the topological charge .
\vspace{-1mm}
\end{itemize}

In particular we hope for applications in the $\epsilon$-regime of QCD
\cite{GasLeu}. In that regime, the pion Compton wave length clearly 
exceeds the box length, $m_{\pi}^{-1} \gg L$, which is an unphysical 
situation. However, simulations in a such small volume may provide
physically significant information, since the low energy constants
of chiral perturbation theory 
take the same
values as in a large volume. Hence the hope is to evaluate them
without the requirement of a large lattice.

However, this implies several conditions for the lattice formulation:
the lattice fermions should keep track of the chiral symmetry and give
access to very light pions. Moreover, the topological charge $Q_{\rm top}$
needs a sound definition, since the measurements are performed at
distinct $\vert Q_{\rm top}\vert $ \cite{LeuSmi}.

Both of these requirements are provided by {\em Ginsparg-Wilson fermions}:
they have an exact, lattice modified chiral symmetry \cite{ML}, and the
fermionic index defines  $ Q_{\rm top} $ \cite{Has}.
Their simulation is now possible, at least quenched \cite{simu}.
Recent results were obtained for the Dirac spectrum \cite{spec}
and for meson correlation functions \cite{corre}.

In particular the Neuberger Dirac operator \cite{Neu1} can be applied.
Its index --- and hence the topological sector --- cannot change
as long as each plaquette variable $U_{P}$ obeys the constraint \cite{HJL,Neu2}
\begin{displaymath}
S_{P} := 1 - \frac{1}{3} {\rm Re ~ Tr} (U_{P}) < \varepsilon
 \simeq 1/ 20.49 \ ,
\end{displaymath}
where $\beta S_{P}$ is the plaquette action.

Therefore $Q_{\rm top}$ is fixed under continuous deformations of the gauge 
configuration if we use the plaquette action $\beta S_{\alpha}$, with
\begin{equation}  \label{Salpha}
S_{\alpha} (U_P ) = \left\{ 
\begin{array}{ccc}
\frac{S_{P}(U_P )}{[1 - S_{P}(U_P ) / \varepsilon ]^{\alpha}} && 
S_P (U_P ) < \varepsilon  \\
+ \infty && {\rm otherwise}
\end{array} \right.
\end{equation}
for $\alpha > 0$. This lattice gauge action (for $\alpha =1$)
was introduced by M.\ L\"{u}scher for conceptual purposes 
\cite{gaugeact}, and applied by Fukaya and Onogi in Schwinger model 
simulations \cite{FuOn}. 

The use of such a lattice gauge action has the advantages that the continuum
property of stable topologies is reproduced, and that tedious computations
of the index can be saved. Finally it allows for the cumulation of
statistics in a specific topological sector; first experience in the
$\epsilon$-regime shows that in particular $|Q_{\rm top}| = 1, \ 2 $
are useful (the sector $Q_{\rm top}=0$ suffers from strong fluctuations,
and at $|Q_{\rm top}| > 2 $ quenched chiral perturbation theory \cite{qXPT}
fails in volumes with box length $L \lsim 1.5~{\rm fm}$).

On the other hand, an obvious problem is that due to
the constraint implemented in $S_{\alpha}(U_{P})$ the physical
lattice spacing tends to be very small. For practical simulations
we hope that $\varepsilon$ values clearly above the theoretical bound 
still suppress topological transitions sufficiently. For instance,
in the Schwinger model $\varepsilon =1$ already stabilized $Q_{\rm top}$
over hundreds of configurations \cite{FuOn}.

\vspace{-3mm}
\section{A LOCAL HYBRID MONTE CARLO ALGORITHM}
\vspace{-1mm}

\begin{figure} \hspace*{-2mm} 
  \includegraphics[angle=270,width=.65\linewidth]{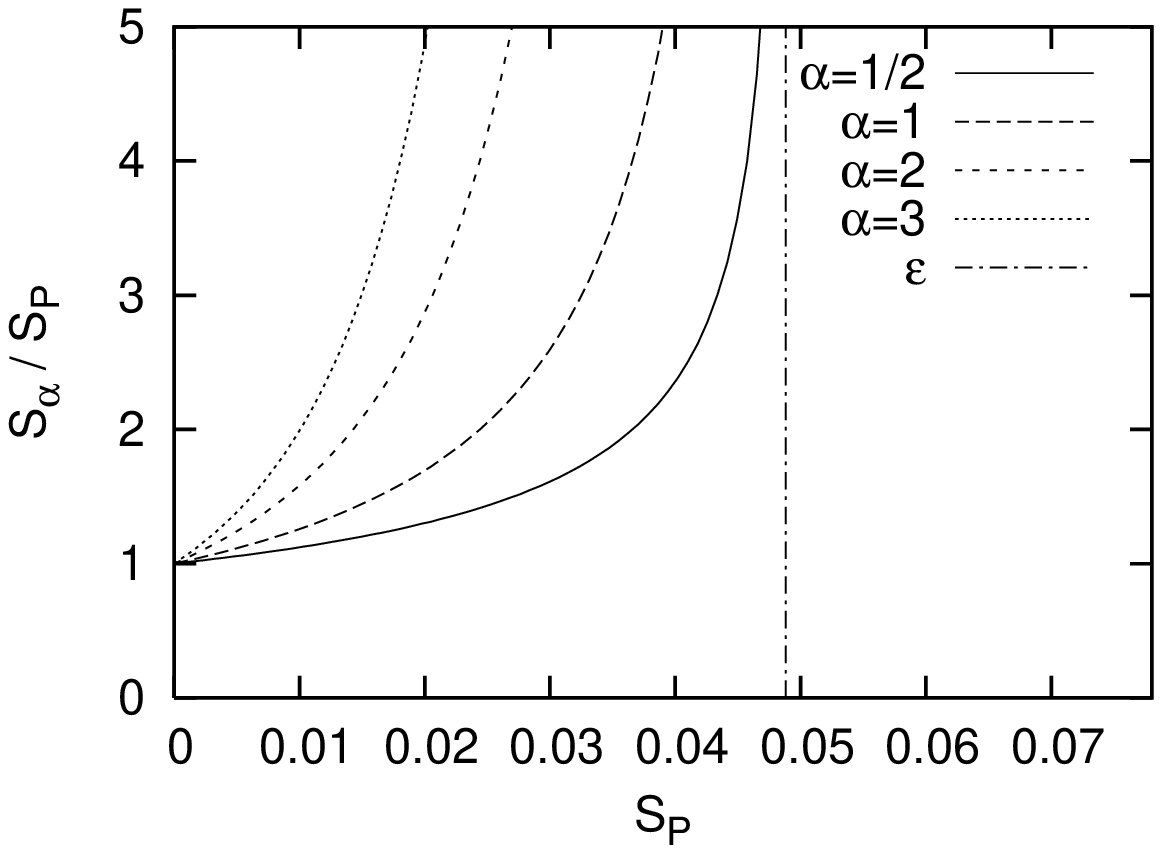} \\ 
  \includegraphics[angle=270,width=.65\linewidth]{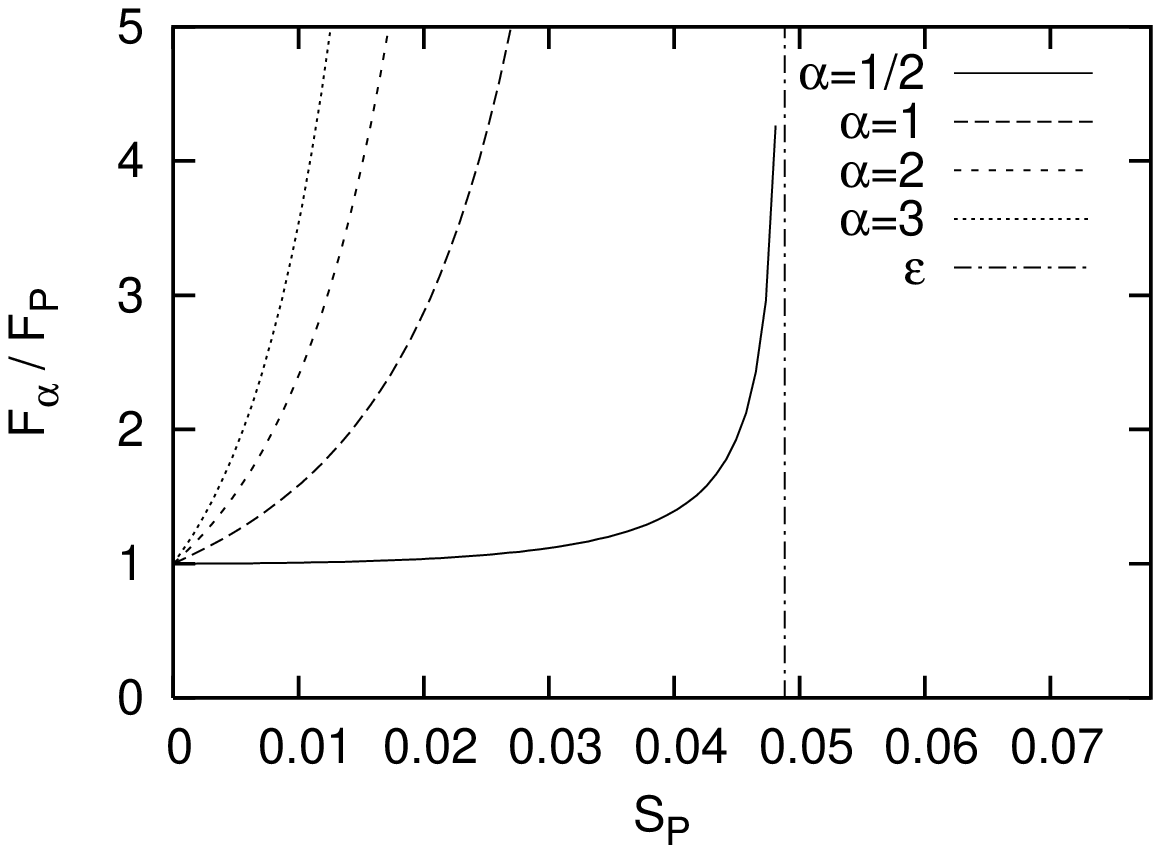}
\vspace{-1cm}
  \caption{{\it Above: ratio between the standard plaquette 
action $S_{P}$ and the modified plaquette action $S_{\alpha}$.
Below: ratio between forces in configuration space, 
which drive the local HMC algorithm.}}
\vspace*{-8mm}
\label{act_for}
\end{figure}

Since gauge actions of the type (\ref{Salpha}) are non-linear in the
link variables, the heat bath algorithm cannot be applied.
Instead we use a local HMC algorithm \cite{lHMC}. Compared to the
Wilson action, the force is just changed by an extra factor on each 
plaquette,
\begin{displaymath}
F_{\alpha} = \frac{\delta S_{\alpha}(U_P )}{\delta U_{x,\mu}} 
= \frac{\delta S_{P}(U_P )}{\delta U_{x,\mu}} \,
\cdot \, \frac{1 + \frac{\alpha -1}{\varepsilon} S_{P}}
{( 1 - S_{P}/\varepsilon )^{\alpha +1}} \ .
\end{displaymath}
The new plaquette action $S_{\alpha}$ and the force $F_{\alpha}$
are illustrated as functions of the standard plaquette action
$S_{\rm P}$ in Fig.\ \ref{act_for}.

\vspace{-3mm}
\section{RESULTS}
\vspace{-1mm}

\begin{figure} 
  \includegraphics[angle=0,width=.75\linewidth]{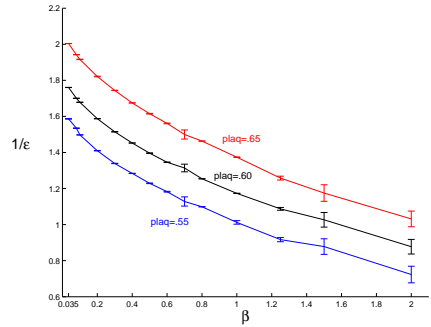}
\vspace{-1cm}
  \caption{{\it The lines of constant plaquette values in the
plane spanned by $1/ \varepsilon$ and $\beta$, 
on a $4^{4}$ lattice.}}
\label{conspla}
\vspace{-7mm}
\end{figure}

We set $\alpha =1$ and as a first experiment we searched for the 
line of a constant plaquette variable $\langle S_P \rangle$ on a $4^{4}$
lattice, as $\beta $ and the action parameter $\varepsilon$ are varied.
The result is shown in Fig.\ \ref{conspla}. As we decrease $\varepsilon$,
very small values of $\beta$
are needed to keep the plaquette constant.

As a more serious approach to identify a line of a constant physical 
scale, we now proceed to a $16^{4}$ lattice and measure $r_{0}/a$
(at $r_{0} = 0.5 ~ {\rm fm}$) following the standard procedure,
see e.g.\ Ref.\ \cite{scale}.
Finite size effects are expected to be on the percent level
for our results presented in Table \ref{r0Tab} (though we only indicate
the statistical error).

\begin{table}
\vspace{-1mm}
\begin{flushright}
\begin{tabular}{|c|c|c|c|c|c|}
\hline
{\footnotesize $1/ \varepsilon$}   & {\footnotesize $\beta$}  & 
{\footnotesize $r_{0}/a$} & {\footnotesize $\beta_{W}$ } 
& {\footnotesize $\tau_{\rm aut}^{\rm top}$} & 
{\footnotesize $\tau_{\rm aut}^{\rm plaq}$} \\
\hline
\hline
{\footnotesize 0} &   {\footnotesize 6.18}  & {\footnotesize 7.14(3)}   
&  {\footnotesize 6.18}   & {\footnotesize ~1.17} & {\footnotesize 7.27} \\
\hline
{\footnotesize 1.25} & {\footnotesize 0.8} & {\footnotesize 7.0(1)~ } &  
{\footnotesize 6.17} & {\footnotesize ~5.76} & {\footnotesize 1.11} \\
\hline
{\footnotesize 1.52} & {\footnotesize 0.3} & {\footnotesize 7.3(4)~} &  
{\footnotesize 6.19} & {\footnotesize 21.04} & {\footnotesize 0.84} \\
\hline
\end{tabular}
\end{flushright}
\vspace{-3mm}
\caption{{\it Results for the physical scale $r_{0}/a$ on a $16^{4}$ lattice
at different values of $1/ \varepsilon$ and $\beta$. 
We add the corresponding $\beta$ value for the
Wilson gauge action, which we denote as $\beta_{W}$. Finally we display
the autocorrelation time with respect to $\vert Q_{\rm top}^{\rm cool}\vert $,
and with respect to the plaquettes, in our MC history.}}
\label{r0Tab}
\vspace{-10mm}
\end{table}

\begin{figure} 
  \includegraphics[angle=270,width=.63\linewidth]{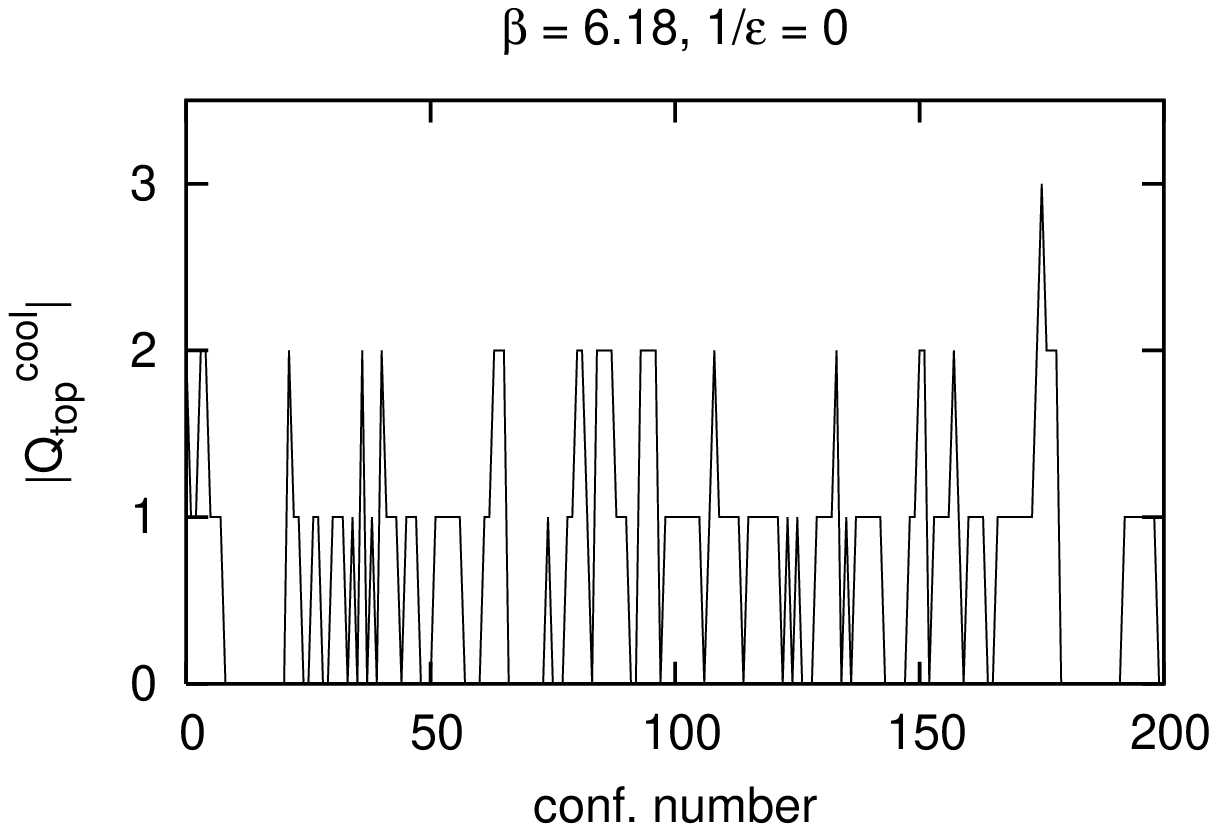} \\
  \includegraphics[angle=270,width=.63\linewidth]{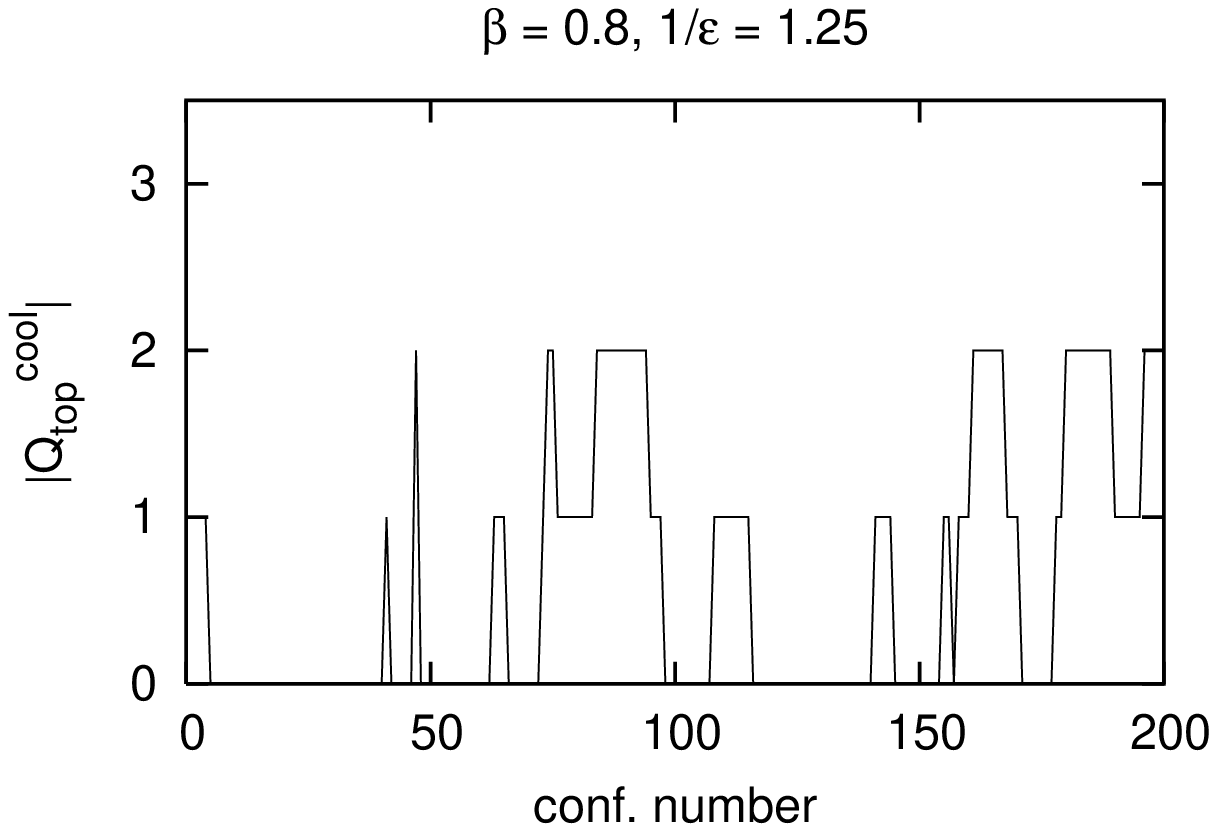} \\
  \includegraphics[angle=270,width=.63\linewidth]{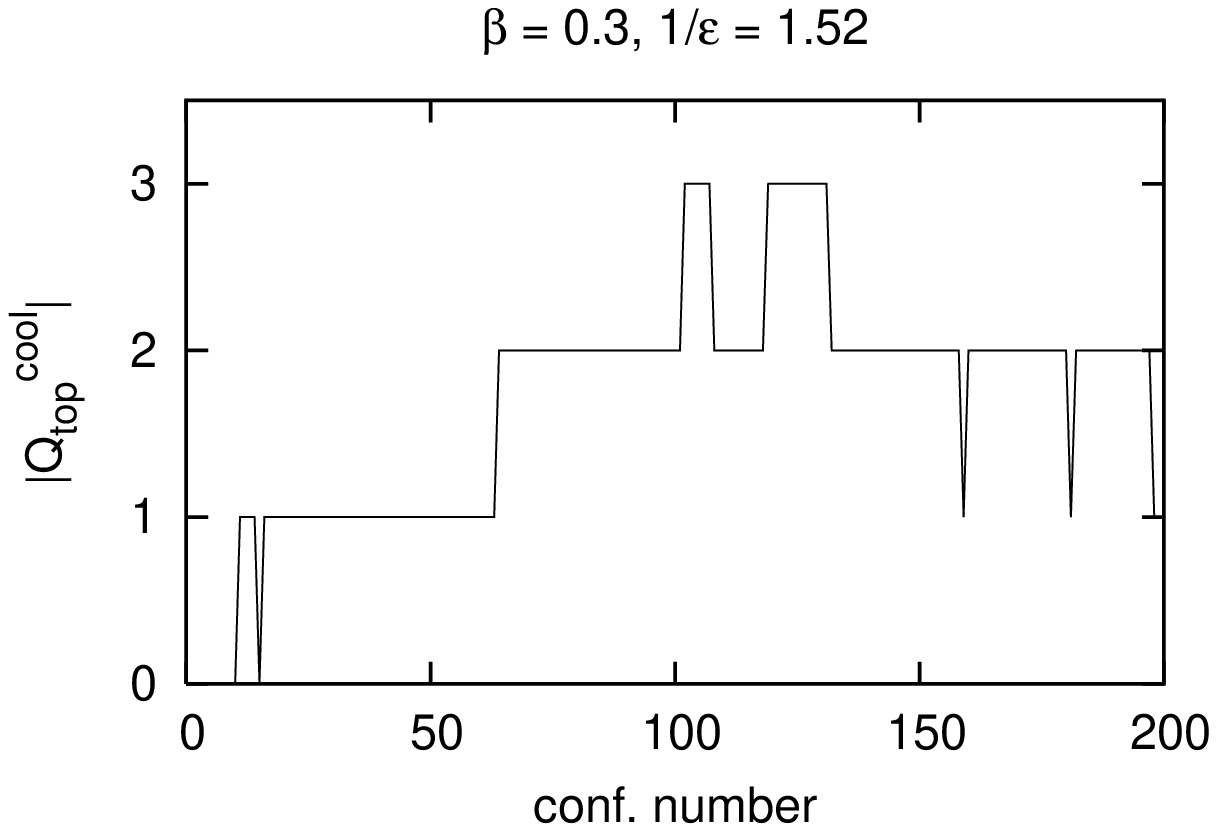}
\vspace{-8mm} 
  \caption{{\it Histories of $|Q_{\rm top}^{\rm cool}|$ on a $16^{4}$ lattice
for the sets of parameters that appear in Table 1. They correspond
approximately to the same physical scale.}}
\label{Qhisto}
\vspace{-8mm} 
\end{figure}

We also performed first tests
of the topological stability. In order to arrive at a quick first
impression this was done with cooling and searching for the first
plateau of the action (which is not sensitive to the sign of 
$Q_{\rm top}$). The histories for $|Q_{\rm top}^{\rm cool}|$ --- ignoring
instanton/anti-instanton cancellations --- are shown in Fig.\
\ref{Qhisto}. 

We recognize a clear progress in view of the topological 
stability as $\varepsilon$ decreases along a line of approximately
constant physics. This can also be quantified by the autocorrelation
in these histories, as the fifth column in Table \ref{r0Tab} shows.

On the other hand, the last column in Table \ref{r0Tab} shows that
the decorrelation with respect to the plaquette values becomes
even better as $\varepsilon$ decreases (at a fixed scale).

\vspace{-3mm}
\section{CONCLUSIONS}
\vspace{-2mm}

``Topology conserving gauge actions''
could be highly profitable in QCD simulations.
The suppression of small plaquette values may speed
up the simulations with dynamical quarks.
A stable $Q_{\rm top}$ is useful
in particular in the $\epsilon$-regime.

We are exploring the applicability of such actions,
in view of the physical scale and the topological stability.
This is an ongoing project.
At present we have first promising candidates for suitable parameters
which provide topological stability to some extent, while keeping
the physical lattice spacing at a reasonable value.

Future tests will involve index measurements to verify the topological
stability. At last we mention that the gauge action (\ref{Salpha})
has the problem that once a plaquette violates the constraint, a force 
in the wrong direction sets in. Hence such configurations had to
be rejected. This happens more frequently as $\varepsilon$ is
decreased. We  now want to test variants of the action (\ref{Salpha}),
where the denominator is replaced by a quadratic or exponential
factor in order to avoid this problem, and
also in view of the caveat pointed out in Ref \cite{MC}.

{\small We thank Martin L\"{u}scher for helpful comments, and
the Deutsche Forschungsgemeinschaft for financial support
through SFB-TR 9-03.}

\end{document}